\documentclass[sigconf,authorversion,nonacm]{acmart}
\AtBeginDocument{%
  \providecommand\BibTeX{{%
    \normalfont B\kern-0.5em{\scshape i\kern-0.25em b}\kern-0.8em\TeX}}}

\copyrightyear{2025} 
\acmYear{2025} 

\setcopyright{acmlicensed}\acmConference[CHIIR '25]{2025 ACM SIGIR Conference on Human Information Interaction and Retrieval}{March 24--28, 2025}{Melbourne, VIC, Australia}
\acmBooktitle{2025 ACM SIGIR Conference on Human Information Interaction and Retrieval (CHIIR '25), March 24--28, 2025, Melbourne, VIC, Australia}
\acmDOI{10.1145/3698204.3716454}
\acmISBN{979-8-4007-1290-6/25/03}

\usepackage{graphicx}
\usepackage{booktabs}
\usepackage{todonotes}
\usepackage[section]{placeins}
\usepackage{comment}
\usepackage{subfig}
\usepackage{enumitem}
\usepackage{multirow}
\usepackage{makecell}

\begin{document}
\title[Understanding Tactics, Trust, Verification, and System Choice in Web Search and Chat Interactions]{Blending Queries and Conversations: Understanding Tactics, Trust, Verification, and System Choice in Web Search and Chat Interactions}
%\title{From Queries and Clicks to Conversations and Blends: How Users Engage with Web and Chat Systems}

\author{Kerstin Mayerhofer}
\affiliation{
  \institution{University of Regensburg}
  \city{Regensburg}
  \country{Germany}}
\email{kerstin.mayerhofer@student.ur.de}

\author{Rob Capra}
\affiliation{
  \institution{University of North Carolina at Chapel Hill}
  \city{Chapel Hill, North Carolina}%{Chapel Hill}
  \country{USA}}%{USA}}
\email{rcapra@unc.edu}

\author{David Elsweiler}
\affiliation{
  \institution{University of Regensburg}
  \city{Regensburg}
  \country{Germany}}
\email{david.elsweiler@ur.de}

\renewcommand{\shortauthors}{Mayerhofer, Capra \& Elsweiler}

\begin{abstract}

%This paper presents a user study (N=22) whereby participants interact with an interface combining web search and a chat function to solve health-related tasks. We employed a concurrent think-aloud protocol to
%capture their thought processes during searches. Qualitative and quantitative analyses of the collected data identified 78 tactics in five categories. Many factors influenced how and why interface features were used, but confidence and trust were important. We show how trust was often misplaced and traded-off against ease-of-use and seemingly perfect-form answers, resulting in overconfidence and incorrect answers.

This paper presents a user study (N=22) where participants used an interface combining Web Search and a Generative AI-Chat feature to solve health-related information tasks. We study how people behaved with the interface, why they behaved in certain ways, and what the outcomes of these behaviours were. A think-aloud protocol captured their thought processes during searches. Our findings suggest that GenAI is neither a search panacea nor a major regression compared to standard Web Search interfaces. Qualitative and quantitative analyses identified 78 tactics across five categories and provided insight into how and why different interface features were used. We find evidence that pre-task confidence and trust both influenced which interface feature was used. In both systems, but particularly when using the chat feature, trust was often misplaced in favour of ease-of-use and seemingly perfect answers, leading to increased confidence post-search despite having incorrect results. We discuss what our findings mean in the context of our defined research questions and outline several open questions for future research.
\end{abstract}

%%
%% The code below is generated by the tool at http://dl.acm.org/ccs.cfm.
%% Please copy and paste the code instead of the example below.
%%


%%
%% Keywords. The author(s) should pick words that accurately describe
%% the work being presented. Separate the keywords with commas.

%\received{20 February 2007}
%\received[revised]{12 March 2009}
%\received[accepted]{5 June 2009}

\maketitle

\section{Introduction}\label{sec:introduction}

Recent developments in Generative Artificial Intelligence (GenAI), along with conversational and search assistants built on this technology, have the potential to support information-seeking tasks in ways that traditional search engines have not achieved \cite{white2024advancing}.

In addition to directly answering user queries, this technology offers several benefits. It can provide an overview of the search space, helping users understand the broader context of their inquiries \cite{liu2024selenite}. It also delivers the potential of personalised responses tailored to the user’s task context, previous interactions, and preferred language style \cite{metzler2021rethinking,murgia2023chatgpt}. The technology can be used to suggest relevant queries to help refine or expand searches and provides synthesised content, such as concise summaries drawn from multiple documents. This could potentially expose users to a wider range of information and reduce the biases they might have when choosing which document to read. % \todo{I read this argument somewhere but cannot re-find it. Not sure we need a cite}. 
Additionally, it can assist with search task management by breaking down complex tasks into more manageable sub-tasks \cite{white2024advancing}.

On the other hand, there are significant problems in using generative AI systems for information-seeking, including issues with hallucination, biases, inaccuracies, lack of transparency, and potential over-reliance on AI-generated content (see e.g., \cite{shah2022situating,shah2024envisioning}).

People are already using generative AI for various information-seeking tasks \cite{choudhury2024exploring}, but little is known about how these interactions affect their behaviours and outcomes. An early exploratory study by Capra and Arguello \cite{capra2023does} examined how information-literate students used an integrated web search and AI chat interface for learning tasks. The study found positive uses of AI chat, such as gaining overviews of unfamiliar topics, getting quick answers without complex subsearches, and generating starting points for web searches. However, despite general distrust, participants often acted on AI chat results (e.g., using them in searches or copying information), suggesting that AI impacts search strategies. "Trust transference" emerged as a concern, where participants trusted unfamiliar AI-provided information when it was presented alongside familiar facts, potentially leading to the acceptance of incorrect information.
Capra and Arguello's results were based on a small, exploratory study involving a homogeneous group of participants performing three complex search tasks. Our goal was to build on this work by gaining a deeper understanding of user behaviour with a similar interface, but using a larger, more diverse sample of participants and a different set of tasks. In the study we present here, we examined how participants performed a series of previously validated health-related information seeking tasks using an interface that combined a GenAI-based chat system with web search capabilities.

Our research questions are:
\begin{itemize}
    \item RQ1: How do people seek information when both GenAI chat and traditional web search options are available?
    \item RQ2: Why do they use these tools in specific ways?
    \item RQ3: What are the outcomes of using these tools?
\end{itemize}

To answer these questions, we conducted a user study with 22 participants, employing a concurrent think-aloud protocol to capture their thought processes during searches. %We analysed their search tactics, identifying 78 tactics in five categories and detailing how frequently each was used.

%[summary of main findings]

%We discuss these findings in the context of Capra and Arguello, and the results of previous studies investigating search behaviour with the same tasks.

\section{Related Work}\label{sec:related_work}

We summarise related work in two subsections: the first explores the practical use of Generative AI systems for information seeking and learning, while the second reviews literature on modelling information-seeking behaviour, which informs our analyses.

\subsection{Generative AI / LLMs for info seeking}

The evolution of search engines has resulted in higher expectations from users, who now anticipate direct answers \cite{stamou2010interpreting} that fulfil their information needs without the need to browse through presented results \cite[]{chuklin2012good, williams2016detecting}. Generative AI systems take this trend even further by allowing for conversational interactions to provide tailored and relevant information to users \cite{metzler2021rethinking}. This development is in line with the early aspirations of the interactive information retrieval community to create more human-like interactions \cite{belkin1984simulation,croft1987i3r}, such as those with a librarian or an appropriate expert \cite{taylor1967question,butler2019health,xu2006will,wildemuth1994information,hertzum2000information}, to resolve information needs. 

The potential of search systems is undeniable, but their pervasive influence also poses significant risks due to their ubiquity \cite{haider2019invisible}. Search engines shape various aspects of our lives, from dietary choices \cite{elsweiler2017exploiting} and health responses \cite{suh2021population,white2009cyberchondria} to how we find professional documents \cite{kruschwitz2017searching}, entertainment \cite{elsweiler2011understanding}, and legal advice \cite{russell2018information}. Despite evidence to the contrary, many people overestimate their ability to assess information credibility. A meta-analysis of 53 studies shows that people often overrate their information literacy \cite{mahmood2016people}, while research indicates individuals correctly answer fewer than half of medical search queries \cite{white2013beliefs}, despite increased confidence after searching \cite{bink2022featured}. Altering search result rankings can change user beliefs \cite{pogacar2017positive}, and featured snippets at the top of results pages can sway attitudes without users realising they've been influenced \cite{bink2022featured,bink2023investigating}. Removing contextual cues, such as the source of the information, as is the case with GenAI systems, could serve to exasperate this problem.

Little is yet known about how GenAI systems are used for information seeking or what impacts they have on search outcomes. Most of the few studies available in the literature are from fields such as education where Generative AI has been shown to 
improve creativity when writing \cite{pellas2023effects}, as well as knowledge gain, self-efficacy and confidence and  motivation while learning \cite{niedbal2023students,songsiengchai2023leveraging,yilmaz2023effect}. 

In addition to the study by Capra and Arguello \cite{capra2023does} mentioned in the introduction, the system Selenite \cite{liu2024selenite} leverages Large Language Models (LLMs) to automatically generate detailed overviews of options and criteria, helping users navigate information more efficiently. Selenite adapts based on user interactions, enhancing both navigation and understanding. User studies show that the system significantly speeds up information processing and improves the sensemaking experience.

Several studies have evaluated the accuracy of tools like ChatGPT in providing health information, revealing that accuracy can vary depending on the topic and factors such as the phrasing of the prompt \cite{koopman2023dr, ayre2024new,razdan2023assessing}. Given these findings, it is likely that people will increasingly turn to such systems for addressing their health information needs. In fact, a recent survey of 611 ChatGPT users found that 44 of them reported using it specifically for health-related queries and decision-making \cite{choudhury2024exploring}. We take this as strong motivation for studying such tasks in this work.

%to read: david \url{https://link.springer.com/chapter/10.1007/978-3-031-61953-3_17}

\subsection{Modelling Search Behaviour}

Search behaviour has been conceptualised at different levels of abstraction. Bates \cite{bates1990should} outlines a hierarchy comprising four levels: moves, tactics, stratagems, and strategies. A \textit{move} is the most fundamental unit, defined as an identifiable thought or action within the search process. A \textit{tactic} consists of one or more moves that are made to further the search. At a higher level, a \textit{stratagem} represents a more complex set of thoughts and actions, incorporating multiple tactics or moves to exploit the file structure of a specific search domain for finding relevant information. Finally, a \textit{strategy} refers to an overarching plan that may include moves, tactics, and stratagems to guide an entire search.

This hierarchy offers a structured approach to understanding and analysing search behaviours at varying levels of complexity. These definitions have served as the foundation for numerous subsequent studies, spanning contexts from professional searchers \cite{fidel1985moves, shute1993knowledge} to web search tasks \cite{savolainen2006user, smith2012internet}, where many of Bates' tactics were confirmed, while others, such as tactics for evaluating search results, were newly identified. Additionally, studies of technical tasks requiring the practical application of information to solve problems have identified further tactics \cite{rutter2019search}. While tactics may be universal \cite{savolainen2017heuristics}, their use is influenced by factors such as the nature of the task, the searcher's knowledge, and the search system in use \cite{xie2012factors}. This raises the question of how well search-based tactics translate to information-seeking interactions with generative AI systems.

While many studies examine individual moves \cite{fidel1985moves, wildemuth2004effects}, our focus, like that of \citet {rutter2019search}, is on tactics, as they offer a model of the cognitive strategies underlying search behaviour \cite{bates1990should, hsieh1993effects}. This emphasis is especially relevant given our use of the think-aloud data collection method (outlined in the following section).

\section{Methodology}\label{sec:method}

To address our research questions, we conducted a user study via Zoom. This setup facilitated participant recruitment and allowed for screen and voice recording. After providing informed consent, participants completed four search tasks using an interface that combined traditional web search and GenAI chat functions, while explaining their actions using a think-aloud protocol. To minimise learning effects, the order of the tasks was varied for each participant. Both interfaces were provided together as we were interested to see how their usage may be combined.

Each task involved answering a health-related question (e.g. ``Do antioxidants help female subfertility?''). For each task, before they searched, participants were asked: 1) their level of familiarity with the task topic (1-7 scale, 7 highest), 2) to give a yes/no answer to the task question, and 3) to provide their level of confidence in their answer (1-7 scale, 7 highest). Then, participants were given the opportunity to seek information using the provided interface (see below). After searching or chatting or both, participants were again asked to provide an answer (yes or no) to the task question and to indicate their level of confidence. 

After piloting the study with five participants, the final study applied the following detailed method.

\subsection{Chat+Search System Design}

\begin{figure*}[ht]
    \centering
    \includegraphics[width=\textwidth]{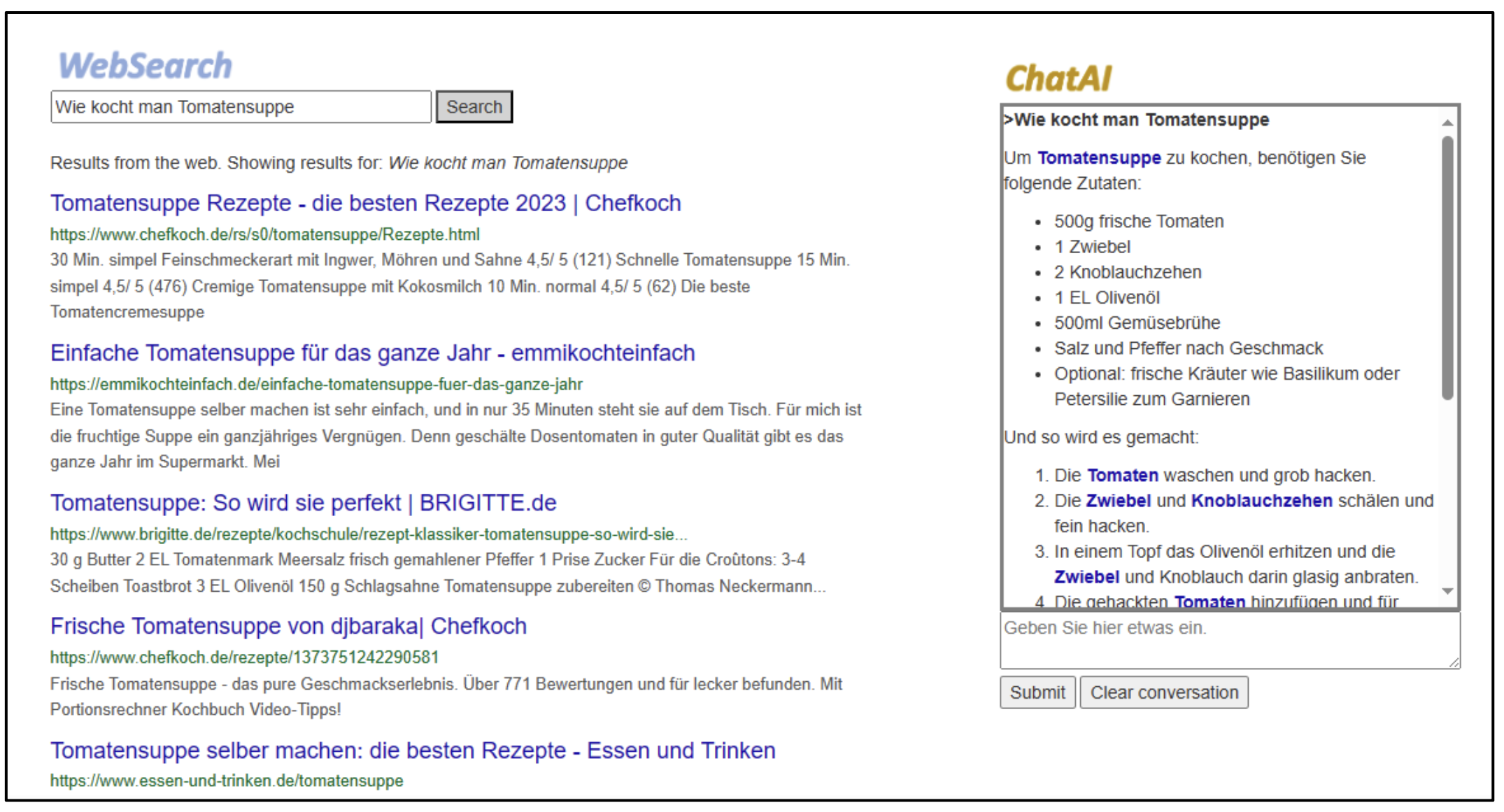}  % Replace with the path to your image
    \caption{Chat+Search Interface}
    \label{fig:interface}  % This label is used to reference the figure in your text
\end{figure*}

%The participants completed search tasks (see Section \ref{sec:tasks}) using a system designed to function exactly like that used in Capra and Arguello's study \cite{capra2023does}.
The participants completed search tasks (see Section \ref{sec:tasks}) using a system designed to function as closely as possible to the one used in \cite{capra2023does}. Our system was implemented using code shared by \citet{capra2023does} and we only made minor changes (e.g., to keep it working with the current version of OpenAI's ChatGPT API). 
The system is illustrated in Figure \ref{fig:interface}.

\textbf{WebSearch:} The left side of the screen (as shown in Figure \ref{fig:interface}) featured a 'traditional' search engine interface labelled 'WebSearch.' This interface included a textbox for entering queries and a display area for listing results. Each result displayed a title, URL, and query-biased text snippet, with results generated using the Bing Web Search v5 API. Users could click on these links to open the corresponding landing pages, just as they would with a typical search engine. Pagination controls (not shown in \ref{fig:interface}) were included at the bottom of the results to allow participants to see additional pages of web search results.

\textbf{ChatAI:} The right side of the screen (also shown in Figure \ref{fig:interface}) contained a chat area labelled 'ChatAI.' This section included a window displaying the chat history and a text area labelled 'Type anything here' (in German, 'Geben Sie hier etwas ein') for users to enter new questions or prompts. When a user entered a prompt, it was added to the conversation history. The entire chat history was then sent to the OpenAI completion API using the gpt-3.5-turbo and gpt-3.5-turbo-instruct models. While waiting for a response, a “Processing...” message was temporarily displayed at bottom of the chat. When the response was returned, it was appended to the end of the chat. 

We used the same setup, parameters and prompts as in Capra and Arguello's work, but had to modify their code to work with the more recent API for gpt-3.5-turbo-instruct.\footnote{Full details of the parameters and prompts can be found at \url{https://osf.io/s37jg/?view_only=d1d7f31d5b25445186a181d154d2db74}} As with their system, to enhance interactivity, we instructed the system to identify key noun phrases from chat responses and convert them into clickable links, enabling users to directly search these phrases through the WebSearch search interface. However, the chat system did not include any direct links or source attributions to the information it returned.
%\todo{@Kerstin, were these used at all? I did not notice any comments about this in the transcripts I read.}

\subsection{Tasks}
\label{sec:tasks}

Participants used the interface to perform a series of health-related tasks based on literature. These tasks were originally developed for a study investigating how search results influence decisions about medical treatment effectiveness \cite{pogacar2017positive} and have since been examined in other interactive IR studies \cite{zimmerman2019privacy,bink2022featured}.

For this study, we used a subset of four tasks used by Bink et al. \cite{bink2022featured}. In their study, \citeauthor{bink2022featured} selected these tasks because they involved topics where the correct answers were not commonly known, leading to greater uncertainty among participants. 
%Details of these topics are provided in Table \ref{tab:topic_choice}. Answers were labelled as correct or incorrect based on the Cochrane review, which is ``a systematic review that synthesises clinical evidence and informs clinical decision-making'' \cite{pogacar2017positive}. A correct answer aligns with the review's findings, while an incorrect answer contradicts them or highlights adverse side effects or harms of the treatment.
Table~\ref{tab:topic_choice} provides details of each task: (1) the task description presented to participants, (2) the efficacy as determined by the Cochrane review (``a systematic review that synthesises clinical evidence and informs clinical decision-making'' \cite{pogacar2017positive}), and (3) the distribution of yes/no/unsure answers given by participants in \cite{bink2022featured}.

\begin{table}[]
\resizebox{\columnwidth}{!}{%
\begin{tabular}{lp{2.75cm}llll}
\toprule
\multirow{2}{*}{\textbf{T}} &
\multirow{2}{*}{\textbf{Medical Treatment}} & \multirow{2}{*}{\textbf{Efficacy}} & \multicolumn{3}{l}{\textbf{Distribution of answers}} \\& & & Yes     & No      & Unsure  \\
\midrule
T1 & Do antioxidants help female subfertility?                & Unhelpful & 14.3\% & 32.1\% & 53.6\% \\
T5 & Do sealants prevent dental decay in the permanent teeth? & Helpful   & 25.0\%   & 14.3\% & 60.7\% \\
T8 & Does melatonin help treat and prevent jet lag?           & Helpful   & 17.9\% & 21.4\% & 60.7\% \\
T10 & Does traction help low back pain?                        & Unhelpful & 25.0\%   & 14.3\% & 60.7\% \\
\toprule
\end{tabular}}

\caption{Medical treatments used throughout the study. Additionally, efficacy labels are provided based on the Cochrane Review. Further, participants answer distributions from \cite{bink2022featured} with respect to the medical treatment are shown.}
\label{tab:topic_choice}
\end{table}

\subsection{Think-aloud-protocol}
\label{sec:thinkaloud}

Concurrent Think-Alouds are a widely used method in the social sciences for data collection \cite{konrad2010lautes}, involving participants verbalising their thoughts while performing tasks, providing insights into their cognitive processes, feelings, and intentions \cite{konrad2010lautes}. This approach allows for immediate verbal reporting of thoughts during task execution \cite{ericsson2017protocol}. The verbalised thoughts are systematically documented and analysed, offering direct insights into the participant's mental processing \cite{konrad2010lautes}.

The transcription followed the simplified method outlined by Dresing and Pehl \cite{dresing2015praxisbuch}, focusing on key information and meaningful participant statements while omitting fillers like "uh" and "um," as well as non-verbal sounds. This approach was sufficient for our needs. Dialects were translated into standard German, and emotional expressions such as laughter were noted in parentheses. Pauses were indicated by starting a new line.

\subsection{Transcription and Analysis}
\label{sec:transcription}

The transcripts and corresponding screen recordings were analysed to identify the tactics and knowledge participants used. The analysis considered participants' overall behaviour, including both their search actions and reflections on their strategies. The transcripts were supplemented with notes on participant actions, using codes from the coding scheme provided by \cite{rutter2019search}. When an appropriate code was not available, we developed our own. As part of this analysis, for each participant + task, we also coded whether the participant used web search only, chat only, or a combination of both during the task.\footnote{This was coded based on observed or self-reported (via the think-aloud protocol) behaviour }%\todo{Todo: need to emphasise that codes encompass not just tactics, but also outcomes, reflections by participants and observations by researchers}

Once a complete set of codes was established, we implemented an iterative process to enhance inter-rater reliability (IRR) and ensure objectivity and reproducibility in the coding. Two individuals independently coded a randomly selected transcript, and IRR was assessed using Krippendorff's Alpha ($\alpha$). Discrepancies were then discussed, and code definitions were refined. This process was repeated until an $\alpha$ value of 0.63 was reached, which was considered sufficiently reliable. The entire set of transcripts was then annotated by a single researcher.

\subsection{Participants}
\label{sec:participants}

Participants were selected using a snowball sampling method, which is cost-effective and leverages personal networks to quickly identify additional participants, enriching the qualitative data collected \cite{biernacki1981snowball}. Seed participants were recruited primarily via student and staff email lists at the University of Regensburg, as well as university fora. The study applied the principle of "theoretical saturation", meaning data collection continued until no new insights were gained, allowing for a focus on analysis and theory development without unnecessary resource expenditure \cite{glaser2017discovery}.

To reach saturation, 22 participants (11 female, 11 male) were recruited. The average age was 29.14 years (SD = 9.62), with ages ranging from 23 to 60. The sample was relatively well-educated, with 40.9\% holding a bachelor's degree, 22.7\% having a master's degree, and 18.2\% possessing a secondary school diploma. Additionally, 4.5\% had vocational training, a specialised high school diploma, a doctorate, or other qualifications.
Seven participants were students, two others were doctoral candidates. The remaining participants had diverse occupations, including one each in unemployment, as an employee, doctor, and optical technician, two each as office workers and educators, one each in semiconductor maintenance, aircraft inspection, psychology, and yoga instruction, and one retiree. 
\section{Results}\label{sec:results}

We structure our results in two main parts. Section \ref{sec:task_diff} presents initial quantitative analyses, showing how task success and confidence varied based on factors like task type, pre-task confidence, and the search interface used. Sections \ref{sec:tactics}-\ref{sec:verification} offer qualitative insights, detailing the tactics applied, user thoughts, and their implications.

\subsection{Task Differences}
\label{sec:task_diff}

\begin{figure}[h]
    \centering    \includegraphics[width=\columnwidth]{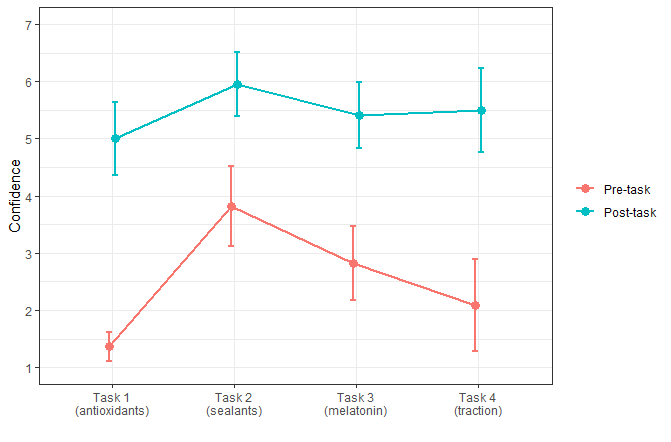} % Adjust the image width to fit the column
    \caption{Pre- and post-search confidence levels for each task. Error bars show 95\% confidence intervals.}
    \label{fig:task_conf}
\end{figure}

\textbf{Familiarity, correctness, confidence:} To start our analyses, we began by investigating the familiarity, correctness, and confidence levels of participants in answering the task questions.
%As noted in Section \ref{sec:tasks}, we used four health-related tasks that Bink et al. \cite{bink2022featured} reported that users had low levels of prior knowledge and confidence about.
As shown in Table~\ref{tab:task_correct}, our participants reported low levels of pre-search \emph{familiarity} with all four of the task topics. The 
\emph{correctness} of their pre-search answers ranged from a high of 77\% to a low of 59\%. Participants' levels of \emph{confidence} in their pre-search answers were quite low on our 7-point scale (Figure~\ref{fig:task_conf}, lower data points).  ANOVA showed a main effect of task on pre-search confidence ($F(3,84)=11.57, p<0.01)$.  Tukey post-hoc tests revealed that Task 1 had significantly lower pre-search confidence than Task 2 ($p<0.01$) and Task 3 ($p<0.01$).  In addition, Task 4 had significantly lower pre-search confidence than Task 2 ($p<0.01$).  These values for pre-search familiarity, correctness and confidence are as we expected -- our goal was to present participants with tasks for which they were not already familiar with the answers.

After searching, post-search confidence increased (versus pre-search) across all four tasks (Figure~\ref{fig:task_conf}). A paired samples t-test found a main effect of pre-vs-post on the level of confidence ($t(87)=-15.95, p<0.001)$. Follow-on tests showed that the post-search confidence ratings were higher than the pre-search (Figure~\ref{fig:task_conf}) for each of the four tasks ($p<0.01$). 

\begin{figure}[h]
    \centering    \includegraphics[width=\columnwidth]{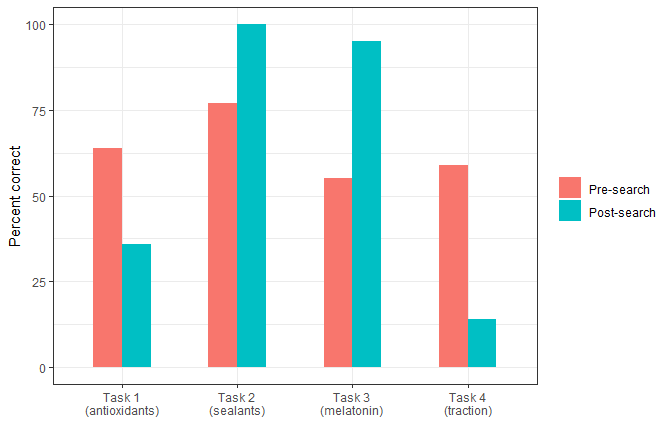} % Adjust the image width to fit the column
    \caption{Pre- and post-search correctness percentages.}
    \label{fig:task_correctness}
\end{figure}

However, post-search \emph{correctness} had mixed results (Figure~\ref{fig:task_correctness}).  For Task 1 (antioxidants) and Task 4 (traction), the percentage of participants giving a correct answer \emph{decreased} from pre- to post-search (e.g., for T1 from 64\% to 36\%).  For Task 2 (sealants) and Task 3 (melatonin), correctness increased from pre-search to post-search.

Considered together, the results in Figure~\ref{fig:task_conf} and Figure~\ref{fig:task_correctness} suggest that for Tasks 1 and 4, participants were not confident about the answer pre-search, but guessed correctly.  Then, after searching, they became more confident, but fewer got the correct answer. This is a concerning result, but is consistent with prior work that has found that searching increases post-task confidence regardless of task success \cite{bink2022featured}. In contrast, for Tasks 2 and 3, participants had higher pre-search confidence and achieved almost perfect correctness post-search. Based on these results, in several of our future analyses, we group Tasks 1 \& 4 and Tasks 2 \& 3.

\begin{table}[]
\resizebox{\columnwidth}{!}{%
\begin{tabular}{llllll}
\toprule
                  & \multicolumn{2}{l}{Before Search} &             & \multicolumn{2}{l}{After Search} \\
\midrule
                  & Familiarity      & \%Correct      & Confidence  & \%Correct      & Confidence      \\
\midrule
T1 - Antioxidants & 1.32 (0.72)      & 64\%           & 1.36 (0.58) & 36\%           & 5.00 (1.45)     \\
T2 - Sealants     & 3.09 (1.60)      & 77\%           & 3.82 (1.59) & 100\%          & 5.96 (1.25)     \\
T3 - Melatonin    & 2.23 (1.80)      & 55\%           & 2.82 (1.47) & 95\%           & 5.41 (1.30)     \\
T4 - Traction     & 1.82 (1.37)      & 59\%           & 2.09 (1.82) & 14\%           & 5.50 (1.66)    \\
\toprule
\end{tabular}%
}
\caption{Summary statistics for correctness and confidence measures for each task.}
\label{tab:task_correct}
\end{table}

\begin{table}[]
\begin{tabular}{llll}
\toprule
& Web search only & AI Chat only & Both \\
T1 - Antioxidants & 27\%     & 14\%      & 59\% \\
T2 - Sealants     & 32\%     & 32\%      & 36\% \\
T3 - Melatonin    & 36\%     & 27\%      & 36\% \\
T4 - Traction     & 23\%     & 5\%       & 73\%  \\ 
\toprule
\end{tabular}
\caption{Task approach.}
\label{tab:task_approach}
\end{table}

\begin{figure}[h]
    \centering    \includegraphics[width=\columnwidth]{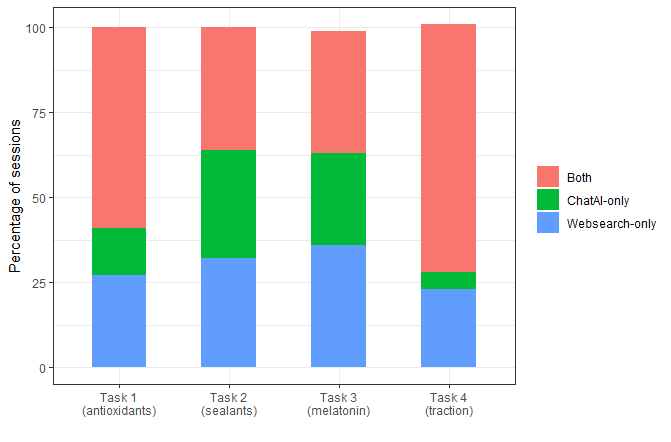} % Adjust the image width to fit the column
    \caption{Percentage of sessions using each approach.}
    \label{fig:task_approach}
\end{figure}

\textbf{Search strategy:} Recall that for each task we coded whether participants used web search only, chat only, or both. Table \ref{tab:task_approach} and Figure~\ref{fig:task_approach} show the percentages of sessions using each search strategy for each task.  Several interesting trends are apparent.  First, Tasks 1 and 4 had lower percentages of chat only participants (14\% and 5\% respectively).  This is also reflected in the higher percentages of participants who used both web search and chat for Tasks 1 and 4 (59\% and 73\% respectively).  Second, we observe that for Tasks 2 and 3 there was a relatively even split among the three options (ranging from 27\% to 36\%). Grouping Tasks 1\&4 and 2\&3, a chi-square test showed that Tasks 1\&4 had a significantly different distribution ($X(2)=9.14, p=0.01)$, with Tasks 1\&4 having lower use of chat-only and more use of both approaches (Figure~\ref{fig:task_approach}).

\textbf{Search strategy and confidence:} The results from Table \ref{tab:task_correct} and Table \ref{tab:task_approach} suggest that when participants had higher familiarity with the topic and higher confidence in their pre-search answers, they were more likely to use chat only.  To investigate this, we conducted an ANOVA to see if there was a difference in pre-search confidence scores given the search approach used (websearch only, chat only, both).  The results showed significant differences ($F(2,85)=3.74, p<0.05)$).  Post-hoc Tukey tests found a significant difference ($p<0.05)$ between chat only ($M=3.47)$ and both ($M=2.2$).  There was also a difference between chat only and websearch ($M=2.46$), but it did not reach statistical significance ($p=0.12)$.  This is an interesting result that suggests that when participants relied only on chat, they had higher levels of pre-task confidence. We did not observe such an effect based on the actual correctness of their pre-search answers.

\begin{figure}[h]
    \centering    \includegraphics[width=\columnwidth]{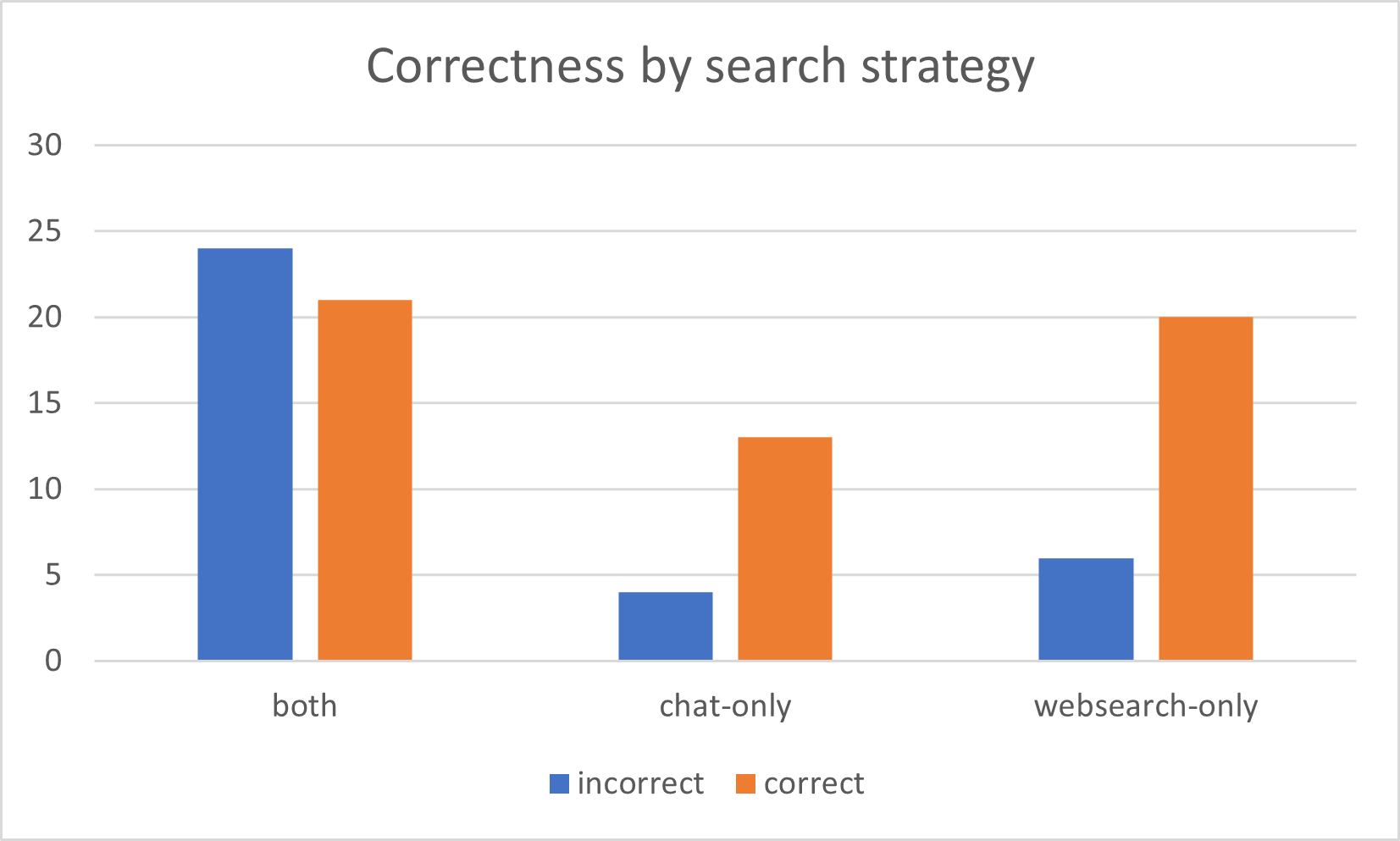} % Adjust the image width to fit the column
    \caption{Number of tasks with post-search answers that were correct/incorrect grouped by search strategy used.}
    \label{fig:correctness_strategy}
\end{figure}

\textbf{Search strategy and correctness:}
Across all three strategies (websearch-only, chat-only, both), we did find a significant relationship between the system used and the probability of obtaining the correct answer post-search, as indicated by a chi-squared test, ($\chi^2(2, N = 88) = 8.39$, $p = 0.015$). Figure~\ref{fig:correctness_strategy} illustrates these differences. Tasks completed using only WebSearch yielded the highest percentage of correct answers (20/26 = 77\%), followed closely by ChatAI (13/17 = 76\%). In contrast, using both systems together resulted in the lowest percentage of correct answers (21/45 = 47\%). This outcome is at least partially explained because participants used both systems more often for the tasks they had the lowest pre-task familiarity and confidence about (T1 and T4).

\subsection{Tactics Used to Solve Tasks}
\label{sec:tactics}

The qualitative analyses of the think-aloud transcripts provides insights into the tactics participants used to complete the tasks and their underlying reasons. In the following sections, we present the findings from these analyses by illustrating how the codes captured user behaviour, supported by examples from the transcripts. Additionally, we include descriptive statistics based on the frequency of these codes. However, to avoid introducing bias, we do not perform statistical tests on this data, as the analysis is exploratory. The patterns discussed here should be viewed as initial observations rather than definitive conclusions. Future experiments with more controlled conditions will be required to establish whether these trends are causal.

\textbf{\emph{How people interacted.}}
We demonstrate that the set of tactics outlined by Rutter and colleagues \cite{rutter2019search} can effectively describe user behaviour when using both the search and chat widgets for health-related tasks. However, not all of the tactics were utilised. In our data, we identified 19 control tactics used for managing information and sources, 24 search and use tactics focused on selecting results and relevant information, and 20 manage tactics related to coordinating the overall task and process. Additionally, we introduced two new categories: 9 reflection tactics, which involve evaluating and trusting search results, and 6 meta-level tactics, which focus on reaching solutions and selecting search formats. Table \ref{tab:tactics_overview} presents the frequency of each tactic type. A comprehensive list of tactics, along with descriptions and examples, is available in the accompanying repository.\footnote{\url{https://osf.io/s37jg/?view_only=d1d7f31d5b25445186a181d154d2db74}}

A total of 78 tactics were identified among the 22 participants. Participants used between 15 and 58 tactics during their searches, with an average of 37.45 tactics. This indicates a generally large repertoire of tactics used by the participants. Three participants (Part 7, 8, and 20) employed fewer than 25 tactics, while four (Part 3, 10, 21, and 22) used more than 45 tactics. 

In this paper, we focus on the tactics that highlight the unique characteristics of chat and search behaviour, as well as how these behaviours are combined. We also explore factors that influence changes in approach and examine aspects related to verification in the context of trust.

\begin{table}[h!]
\centering
\resizebox{\columnwidth}{!}{%
\begin{tabular}{|l|c|c|c|}
\hline
\textbf{Tactic Type} & \makecell{\textbf{Num. Identified} \\ \textbf{Tactics}} & \makecell{\textbf{Num. Tactics used} \\ \textbf{by Parts (N = 22)}} & \makecell{\textbf{Avg. Num. Tactics} \\ \textbf{used per Part}} \\ \hline
Control Tactics & 19 & 637 & 28.95 \\ \hline
Search and Use & 24 & 434 & 19.73 \\ \hline
Manage & 20 & 533 & 24.23 \\ \hline
Reflection & 9 & 754 & 34.27 \\ \hline
Meta-level & 6 & 391 & 17.78 \\ \hline
\textbf{Total} & \textbf{78} & \textbf{2749} & \textbf{124.95} \\ \hline
\end{tabular}%
}
\caption{Overview of identified tactics: Number per Particpant/Overall/Average}
\label{tab:tactics_overview}
\end{table}

%The analysis of coded behaviour highlighted two main tactics users employed to communicate their needs: (C7 Exhaust) was assigned to both WebSearch and Chat AI interactions when the participants submitted the entire question to the appropriate interface. This approach was often paired with the use of available aids (C4 Support). User 20 explained, "Let's make this easier," before copy/pasting the entire question into the search bar.
%On the other hand, Keyword or Single Term Input (C6 Reduce) was more common in WebSearch (e.g.'Traction for pain’, Part. 2 (web search). 

\subsection{Tactics with WebSearch and Chat}\label{sec:tactics_web_chat}

Different participants employed varying approaches depending on whether they were using WebSearch or ChatAI. For example, two participants explicitly stated that they preferred entering full questions in ChatAI, while using keywords in WebSearch. Additionally, seven participants in 12 instances did not explicitly describe this tactic but followed a similar pattern, using keywords for web searches and full questions for chat searches.
"For WebSearch, I wouldn't directly enter a question, but rather use individual search terms." (Part 3, section 34). 
 One person used a tactic of initially entering all relevant terms and then narrowing down the search further: "Here, I just do Ctrl C, Ctrl V, and then I always look for more precise terms. The longer the answer, the more I refine the query. If I don't get a satisfactory result, I keep narrowing it down to what's relevant—that's my approach." (Part 18, section 30).

Copy/pasting was not the only (C4 Support) function. Participants also used the GenAI’s ability to translate text to change the language to German or English, with one participant noting, "I'll switch it to German and read it quickly" (Part 3, Section 174). Another commonly used function was the shortcut CTRL + F for finding words on a website. For instance, Participant 3 employed CTRL + F to search for "permanent," and Participant 12 used it to find "sealing," commenting, "I'll search for the word sealing on this page to shorten things [task times]. The word appears 4 times" (part 12, Section 86 - 87). This was a means to quickly determine relevance, i.e. whether the page did actually have information on the topics they were looking for. We did not find any examples of participants using the chat feature to generate query terms, as been suggested in the literature \cite{white2024advancing}.

%Additionally, one participant tried using advanced search features like "+" in the web search, though this was not supported by the system. These tools were likely used to enhance search efficiency and save time.

Query reformulation (C12 vary) was used to optimise web search results. For example, Participant 6 changed an initial query to "traktion schmerzen lws [German]" in the web search interface, explaining, "I used that term because it’s an abbreviation for lower back pain" (Part 6, Section 115 - 116). Similarly, Participant 12 mentioned, "I would actually replace a few words, like 'antioxidants' and 'woman,' to see if I can find something" (Part 12, Section 44). In contrast, the chat interface primarily received entirely new questions to address.

Certain behaviours were observed less frequently in ChatAI compared to WebSearch. %These include the tactics M1 (Verification) and SU25 (Consensus). The lower use of M1 (Verification) with Chat AI might be because users tended to accept the results without questioning them. \todo{need numbers on this. TODO: David. 
%For example, the code SU25 (Consensus), which involves comparing information from multiple sources to reach a consensus or compromise, was
%only applied in WebSearch. The fact that Chat AI responses typically combined various definitions and aspects of answers, eliminated the need for participants to seek additional sources for comparison. 
%\todo{DE: add concrete numbers here if possible}
The M22 (Select) tactic, for example, which involves breaking down a task into subtasks and working on them individually, was more common in WebSearch. The chat often eliminated the need for this tactic as the responses typically combined the results of the separate subtasks as they were performed using the WebSearch tool. Conversely, the (SU1 Visual Clarity) tactic was observed more frequently with ChatAI than with WebSearch. This code refers to the selection and use of an object because its presentation and format make it easier to use  (e.g. "Aha, that is well explained.", Part 1). This was because ChatAI often provides a single, nicely packaged answer. %Additionally, (self-criticism) was more commonly seen with ChatAI, likely because Participants were aware that they were trading-off ease of use and less effort on their part against the potential for misinformation or the absence of source citations. For example,
%"Of course, I haven't read any studies now, but yes." (Part 5) and "But I didn't read carefully enough there. [but I will accept the answer]" (Part 3). 
%Generally, as explained in Section \ref{sec:verification}, the method and extent of verification differed depending on whether ChatAI or WebSearch was used.

\subsection{Changing Approach}

In 30 cases, changing tactics involved modifying or adjusting an existing approach, strategy, or method (M13 Change Task). This was often done when the current tactic was not yielding the desired results. 
For example, "What I was searching for didn’t help, so I’d probably look up 'caries and sealings'" (Part 3, Section 263 - 265). Participant 13 noted, "This didn’t seem worthwhile since it’s more related to vehicle technology." (Part 13, Section 105). %User 12 remarked, "Since the commands that work on Google don’t work here, I’d directly Google the question to see if there’s any information" (User 12, Section 38).
This tactic was also applied when circumstances or requirements changed, allowing participants to better respond to challenges or find more effective ways to achieve their goals. Participant 13 further commented, "I’d rather go back and continue working with the right side (ChatAI), as reading everything seems too much" (Part 13, Section 44).
These findings suggest that people adapt their approach or tactics when they find (or want to find) better, easier, or more efficient methods to achieve their goals or desired results. Changes in tactics often occur if users realise that current strategy is ineffective or that more efficient alternatives are available. This reflects an adaptive mindset and the ability to adjust to changing conditions for greater success.

Participants typically switched from the search interface to the chat function to improve search effectiveness when results were unsatisfactory. Eight participants switched from web to chat twelve times when their searches did not yield the outcomes they had hoped for. Participant 1 remarked, "Honestly, I’m not finding anything. I might try the ChatAI instead," before entering a query into ChatAI (Part 1, Section 62-67). Similarly, Participant 17 noted, "I find this unsatisfactory. Since I have the chatbot, I'll just copy this question into it," and did so (Part 17, Section 157-159). Participant 21 also switched, observing, "I'm not getting much information. The ChatAI might be useful here" (Part 21, Section 286-287).

Conversely, participants did not tend switch from ChatAI to WebSearch out of dissatisfaction with the results—the chat interface always provided responses that answered the participant's question. However, some participants returned to WebSearch to verify the information because they were uncertain about the trustworthiness of the chat-generated answers. In the next section we examine verification or lack there of in more detail. % For instance, participant 11 said, "So, I want to check what Chat AI said" (participant 11, Section 137). participant 15 expressed scepticism, stating, "Since AI is known to give misinformation, let's check this on the website as well" (User 15, Section 31). User 14 added, "Given that chatbots sometimes make things up, I'll double-check on the website by entering the full query again" (User 14, Section 73). \todo{need numbers to back this up because it does not fit with the }

\subsection{Verification and Trust}\label{sec:verification}

We observed differing approaches to validation, ranging from detailed searching to validate sources to no validation at all. Below we summarise what our data inform us about validation and how this related to whether WebSearch or ChatAI was used. 

%\todo{David: we should look at the relationship between pre-task confidence in answer and whether they verify. Also it would be good to look at the relationship between validation and task success.}

Three important tactics for verification were SU24 Discriminate, SU25 consensus, and SU27 internal verification. 
SU24 Discriminate was applied when the participant compared different information sources to determine which is superior. Examples from our data include: "That article is much more recent." (Part 16, section 172 - 180) and "Researcher: Aha, so you don't see Liebscher and Bracht as a good source?" Participant:  “No (laughs), I like Springer better" (Part 6, section 117 - 118).

SU25 consensus was applied when information from multiple sources was compared in order to find a consensus. Often, this involved some kind of compromise in order to combine multiple information fragments into a single outcome. In our data, examples include: "So, with the feedback from the AI and the video [found via WebSearch], I now have enough to answer the question.” (Part 5, section 103); “Two sources say that it’s very likely to help. Okay. With that, I’ll say the search is finished for now.” (Part 11, section 51); “And since the last responses that ChatAI gave me pretty much matched what I had found elsewhere.” (Part 11, section 123)"

SU27 internal verification involved comparing found information with information that the participant  already knew or believed. Sometimes this meant verifying prior knowledge e.g.
"Oh look, I was right. It depends on the antioxidants, it's [the task] just really hard." (Part 6, section 171);
"That already confirms it for me." (Part 16, section 147 - 148); "Along with what I already know from my own experience, that melatonin is either not dangerous or is safe, I would say, and if I were now planning a trip or something like that, I would assume that I can use it because it’s safe." (Part 12, section 138).

%Other times it involved corroborating what they had found in earlier interactions e.g. "Das steht jetzt wieder genau das Selbe." (Part 15, Absatz 185)

This code was also applied when new information was found that showed something different, but complementary e.g. "... this article doesn't seem to address its actual effectiveness, but only talks about how it might help, explains the risks, how it's done, and when the best times for it are." (Part 12, section 81).

In total, 14 people paid attention to sources or checked where specific articles originated (SU22 Trust). They looked at the domain or source of the articles and chose certain articles based on their origin. Examples include: "Science. Cochrane. Yes, that would also be a good source." (Part 6, Section 44), "Ah, look. There it is, the answer we want. That also looks like it’s from a reasonable source." (Part 10, Section 44); "And right now, I'm really looking at the sources. So, where certain websites are at the top with green marks, whether something looks familiar, from studies I know, that are relatively reliable." (Part 11, section 38).

%[note SU22b only headline] "Interviewer: But why the first two? Now because of. Part 11: Because of the headlines. They just kind of spoke to me. So I think one was "causes treatment and comparison" and the second was "how is it diagnosed and treated"." (Part 11, section 106 - 107)

The code (SU22 Trust) was only applied to WebSearch interactions since the source / origin of information was not provided by the chat by default. None of the participants asked the chat for sources. When evaluating sources (SU22 Trust), seven participants considered how old the information was (SU15 Current). This influenced their search behaviour, as older publication dates led them to avoid or view these articles more sceptically.
"Published in 2017, that's too old for me." (Part 1, Section 57); "2002 might not be the most recent study." (Part 6, Section 75); "Hmm, I see that this article is from 2013, which is already over 10 years ago." (Part 14, Section 129). Newer articles generally received more attention, and the information they contained were typically preferred. "Okay. Serious risk of bias. Okay, this is a meta-study from 2020. Well, alright." (Part 10, Section 205)
- "(Reads title), oh, 2021, that's nice." (Part 6, Section 168)
- "Yeah. I want to quickly check. This article is much newer." (Part 16, Section 180).

\begin{figure}[h]
    \centering
    \includegraphics[width=\columnwidth]{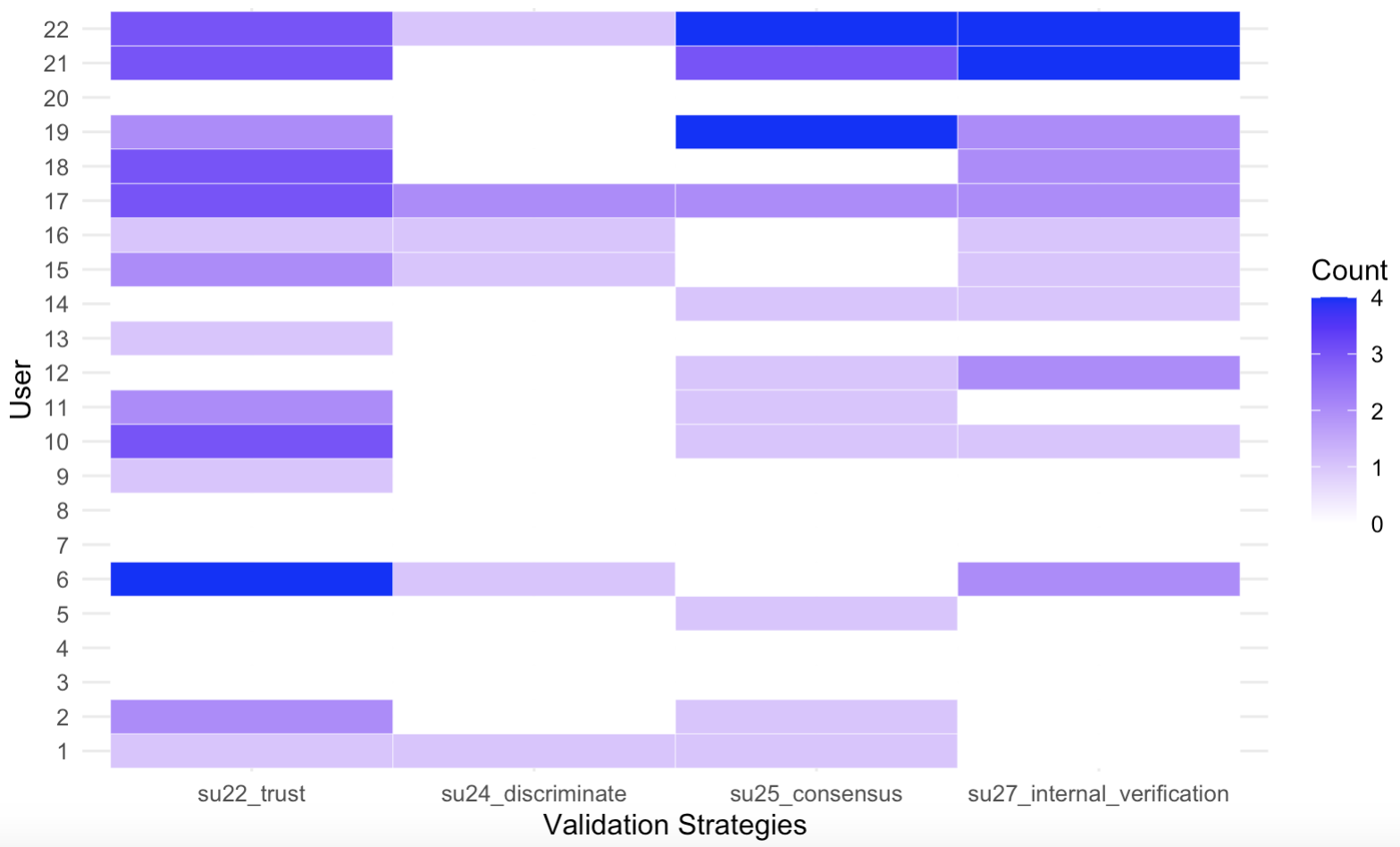} % Adjust the image width to fit the column
    \caption{Number of Tasks using each Validation Strategy per Participant}
    \label{fig:verification_strategies}
\end{figure}

Only one participant attempted to validate a chat response by means of the chat itself asking the ChatAI system for studies evidencing the claim. 
 "That's why I'm asking the chatbot again which studies prove that antioxidants help, and then I would check if they are more recent. Maybe research has changed or discovered something new.
Participant 14 types 'Which studies say that antioxidants help with female subfertility?' into ChatAI." (Part 14, sections 129 - 131). This did indeed lead to the participant having the correct answer.

Figure \ref{fig:verification_strategies} shows the number of tasks for which each participant utilised the key verification tactics. This illustrates the extent to which focus on verification varied across participants. Some participants, such as participants 6 and 22 considered this in every task, whereas participants 3, 4, 7, 8 and 20 made use of no verification tactics at all. Some participants who used the chat  relied entirely on the responses provided by the system. The reflection code (trust reflection) provides some insight their thought processes in relation to this: "Hmm. The only thing that doesn't quite fit or what might bother me a little is that there is no source. I don't know if what is written there is true or not. Unfortunately, I can't verify whether it's correct (laughs). But no, I would actually trust 100\% what the right-side ChatAI says." (Part 13, Section 53).

Participant 4, who ended up with the incorrect answer to the traction task using the chat interface, felt he got a more precise answer to what he was looking for with the chat, whereas his WebSearch gave him more general background information. "Yes, exactly. I read the chat text and it gave me a pretty clear answer to the question, answered it, although with the web search, it's mostly about the causes and treatment of back pain and doesn't really go into traction." (Part 4, Section 53)

Participant 9 provided similar comments. After completing earlier tasks using both web search and chat, and struggling to piece different parts of the task together, i.e. definitions for concepts relevant to the task, he commented on the ease of use: "And then I enter it back into the chat. I've kind of taken a liking to it now (laughs) because it's really easy and actually user-friendly. Somehow, I'm noticing the difference between a regular Google search, or web search as it's called here, and ChatGPT or the AI itself. It really gives concise, insightful answers to a question that are solid and make sense" (Part 9, section 98). This participant also explicitly commented on the trustworthiness of the answers he achieved. "What I’ve noticed is that with the AI, you have much more reliable answers.” (Part 9, Section 123). Unfortunately, this trust was misplaced in this case as the participant answered the task incorrectly.

The ease with which answers could be attained was commented on regularly: %"That was too much text for me. I'll quickly check for traction to see if it's mentioned somewhere." (Part 11, Section 74);
"[I would] rather go back and continue working with the right side (ChatAI), as reading everything seems too much" (Part 13, Section 44).  There was also evidence that the ease of use seemed to outweigh a trust with some participants simply accepting an answer from the chat since it was in the right format and sounded plausible: "That actually sounds logical, so with that, it's basically already answered." (Part 2, Section 161)
%“Hört sich eigentlich logisch an, also nach dem ist das eigentlich schon beantwortet.” (User 2, section 161)

Participants reported being  aware that they were trading-off ease of use and less effort on their part against the potential for misinformation or the absence of source citations. For example, "Of course, I haven't read any studies now, but yes." (Part 5) and "But I didn't read carefully enough there. [but I will accept the answer]" (Part 3). The code 'self-criticism' was applied to such examples.

Other participants mentioned doubts about the trustworthiness of chat interface: "But since we know that chatbots sometimes make things up, I’ll check the website again and enter the whole question there as well." (Part 14, Section 73); “Use it, then question ChatGPT again, because you never know where it pulls its data from.” (Part 22, Section 113).

%“User 22: Okay, I would first let Chat AI find the answer and then check it again with Google.” (User 22, section 27)

%“User 22: Einsetzen, nochmal hinterfragen Chat GPT, weil man weiß ja nie, wo der die Daten herzieht.” (User 22, section 113)

%“User 22: Okay, ich würde jetzt erst mal Chat AI die Antwort herausfinden lassen und die dann nochmal überprüfen, mit Google.” (User 22, section 27)

\begin{figure}[h]
    \centering
    \includegraphics[width=\columnwidth]{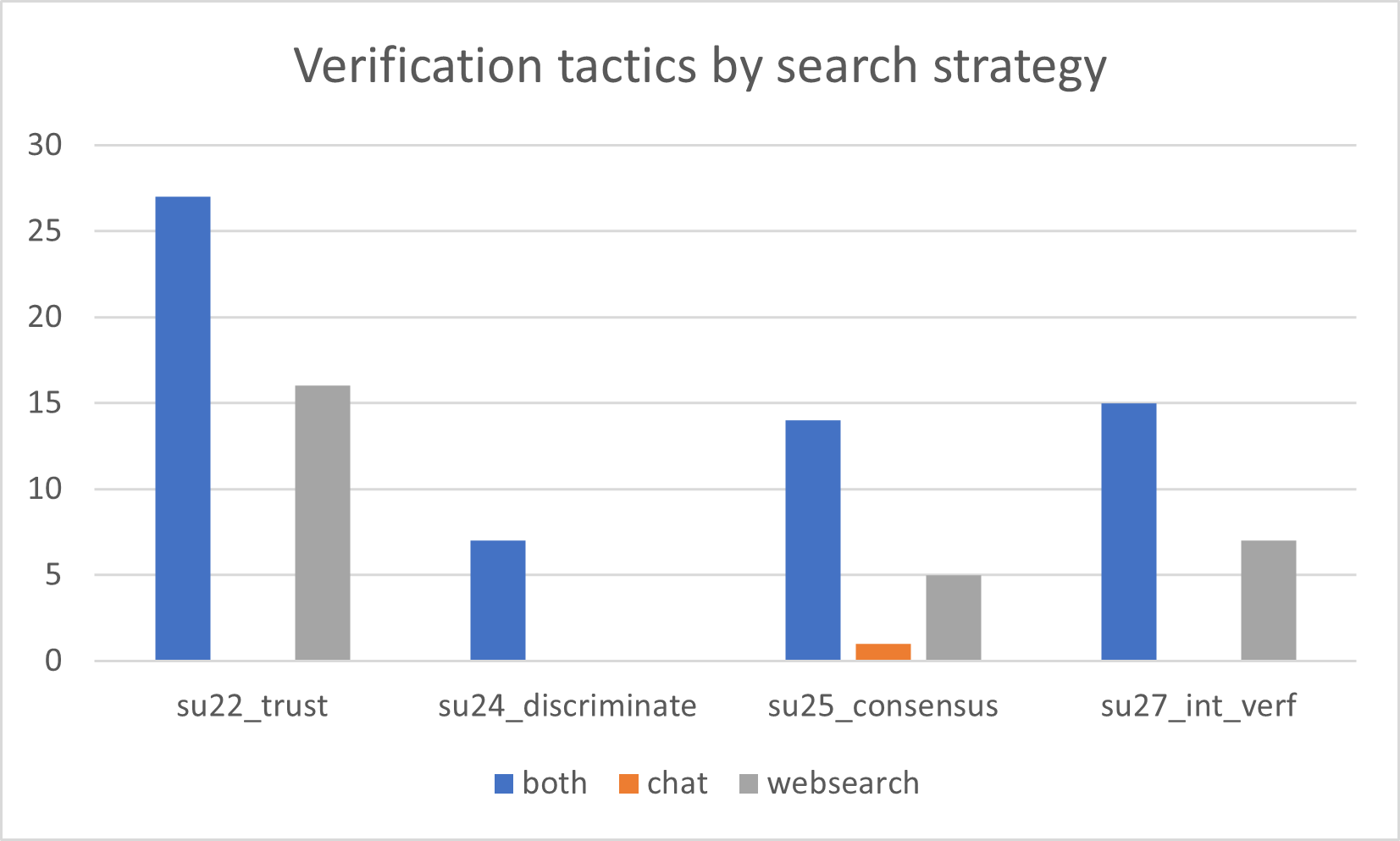} % Adjust the image width to fit the column
    \caption{Number of Tasks using each Validation Tactic across difference search strategies}
    \label{fig:verf_tactics_strategy}
\end{figure}

The verification tactics employed differed depending on the interface widget being used. Table~\ref{tab:verf_tactics_strategy} and Figure~\ref{fig:verf_tactics_strategy} show the number of tasks for which a particular tactic was used when the participant used a particular search strategy (websearch-only, chat-only, both). There are several interesting points to observe.  First, almost no verification tactics were used when the participant chose a chat-only strategy.  In only one chat-only case did a participant use the SU25 Consensus strategy. This was the successful example described above where the participant asked the chat for studies evidencing the answer. Second, verification tactics were used in more of the tasks when participants used both web search and chat. One reason for this could be that tasks where both were used tended to be the more difficult tasks. Another reason is that participants may have used one of the two systems (web search after chat) to verify the other. Finally, we note that the SU22 Trust tactic was the most commonly used verification method. This could be because it was the most superficial and required the least effort, as it involved only making a subjective judgment rather than verifying information against other sources.

\begin{table}[]
\resizebox{\columnwidth}{!}{%
\begin{tabular}{llllll}
\toprule
        Approach & su22\_trust & su24\_discriminate & su25\_consensus & su27\_int\_verf \\ \hline
        both & 27 & 7 & 14 & 15 \\ \hline
        chat & 0 & 0 & 1 & 0 \\ \hline
        websearch & 16 & 0 & 5 & 7 \\ \hline
    \end{tabular}%
}
\caption{Number of tasks using each validation tactic across different search strategies.}
\label{tab:verf_tactics_strategy}
\end{table}

Despite the variety of verification tactics used, we did not find a significant correlation between using a verification tactic and answering correctly post-search.  Across the 88 search sessions (22 participants x 4 tasks), the Pearson correlation between using at least one of the verification tactics (SU22, SU24, SU25, or SU27) and answering correctly was only 0.03 -- a negligible correlation.

\section{Discussion}

In this study, we explored how Generative AI systems, in the form of chat interfaces similar to ChatGPT, can be used alongside traditional web search engines in health-related information-seeking tasks. Our findings show that GenAI systems are neither a perfect solution to all issues nor a significant decline from standard web search performance. 

%\subsection{Research Questions}
We explored three main research questions: (RQ1) How do people seek information when both ChatAI and web search are available? (RQ2) Why do they use these tools in specific ways? and (RQ3) What are the outcomes of using these tools?

In terms of RQ1, we found several interesting results.  First, participants used a wide range of tactics during their searches.  We found that the set of tactics outlined by \citet{rutter2019search} worked well to capture search behaviours.  However, some of the tactics were not observed (or observed less much frequently) in the chat behaviours (e.g., verification, query reformulation, consensus).  Second, we observed that participants used a mixture of high-level approaches, including tasks where they used WebSearch only, chat only, and both components. This illustrates that participants were able to quickly start using the ChatAI component, make decisions about when to incorporate it (or not) into their searches, and see ways to combine its use with traditional web search.
%We observed a wide variety of interaction patterns with both tools. Tactics differed between GenAI-chat and web search. For instance, while web search queries consisted of keywords, ChatAI input tended to be full questions. 
Third, we observed some creative use of ChatAI, such as for translating information between languages. However, we did not observe the use of ChatAI for navigating search spaces or generating queries, which has been suggested in previous studies e.g. \cite{white2024advancing}. Future research should explore whether these behaviours could enhance search effectiveness and how they could be encouraged through nudge or boost interventions \cite{ortloff2021effect,bink2024balancing}. White's \cite{white2024advancing} claim that AI assistants could eliminate the need for task management behaviours was supported in our case. When ChatAI was used, it often handled separate sub-tasks, such as looking up definitions for key concepts on behalf of the participant. Participants appreciated this and took advantage of it!

In RQ2, we considered why participants used the tools in specific ways. Again, we found several interesting results. %First, we observed that many participants used full questions with chat versus keywords with web search.  GenAI systems have demonstrated their ability to respond to long-text prompts, and our participants seemed to take advantage of this.  Second, we noted that web searches had more query reformulation, but in the ChatAI component participants mostly asked new questions rather than reformulating previous ones.  This is an interesting dynamic that could be an artefact of our task types.  Future work should explore this aspect. 
The results show that the nature of the task and how it was perceived played a critical role in whether participants relied on search or chat. For tasks where users had high confidence pre-task, the chat system was used more often. This result resonates with the finding from ~\citet{capra2023does} that users' level of familiarity with a topic influenced their trust (or lack of trust) in AI chat.  However, it differs from their finding that people made use of AI Chat for unfamiliar topics~\cite{capra2023does}.  This may have been due to the more open-ended nature of their tasks compared to our tasks which were more focused on finding a correct answer to a question.

User experience with the technology also seemed to impact the choice of system used. For example, some participants mentioned recognising the strengths of ChatAI in early tasks, leading them to use it more prominently in later ones. The ease of use of the system and the seemingly plausible answers it provided were highly appealing to participants. This mirrors comments from users' in ~\cite{capra2023does} who reported that GenAI chat provided concise, easy-to-understand answers that often ``sounded reasonable''.  While it is encouraging that users took advantage of ChatAI's abilities to provide clear and easy-to-read answers, this also points to risks that users may accept GenAI hallucinations as true just because they are presented clearly.

Furthermore, we observed that participants tended to switch from web search to chat to improve the quality or relevance of the information retrieved.  However, switches from chat to web search were mostly to verify information rather than due to non-relevant results being returned from the chat.  This is also an interesting dynamic that bears additional exploration.  Although we observed many cases where participants accepted chat information without verification, it is encouraging that web search was used to verify it in some cases. The verification tactics were not always successful and it is obvious that users need support with this.

RQ3 explored the outcomes of using the web search and ChatAI tools for the assigned tasks. While some scholars suggest that AI chat systems may introduce new challenges, we found no evidence of a significant decline in task performance compared to using web search alone. Overall, our participants achieved a post-search correctness rate of 61.4\%, slightly lower than the 65\% reported by Bink and colleagues for the same tasks using a standard web search engine \cite{bink2022featured}. In our study, for the two tasks where participants were unfamiliar with the topic and had low confidence in their initial answers (T2 \& T3), they were largely successful in finding the correct answer, regardless of the system or approach used. For the other two, more challenging tasks (T1 \& T4), the choice of system also had little impact on outcomes, with overall performance being poor (34\% and 14\%, for tasks T1 \& T4, respectively). What is especially concerning is not only that participants performed very poorly on these tasks, but that participant confidence in their answer often increased during searches, regardless of whether or not it was correct. This highlights the need to improve both the systems and user behaviours to address these issues.

Verification strategies varied significantly across participants, and depended on whether chat or search systems were used. Surprisingly, we found no evidence that more extensive verification increased the likelihood of reaching the correct answer, particularly for the more difficult tasks. For example, in Task 1 and Task 3, where many participants failed to arrive at the correct answer according to the Cochrane Institute’s consensus, verification efforts did not lead to better outcomes. This result is concerning, but it aligns with previous research showing the diversity and unreliability of credibility judgements \cite{kattenbeck2019understanding}. It also reinforces the findings of recent work showing that searching online to evaluate misinformation can be problematic and often leads to misinformation being more strongly believed \cite{aslett2024online}. One observed verification tactic, which was successful in the single instance it was used, involved asking the chat for studies to support the claim made in its response. An open question remains as to whether this approach is generally effective, warranting further research.

\section{Conclusions, Limitations and Future Work} \label{sec:conclusion}

The quantitative and qualitative analyses in our study shed light on how people behave when Generative AI (GenAI) chat is integrated into web search, highlighting both the potential and challenges of this approach. Our findings indicate that while the integration has the potential to enhance information-seeking, there are key limitations and considerations that must be addressed.

It is important to note that this was a small-scale study with a relatively homogenous set of tasks. Although the study reached a point of saturation with 22 participants, meaning we ceased to observe new behaviours beyond Participant 19\footnote{Saturation criteria typically involve stopping after 3 participants with no new behaviours.}, it remains possible that different or more complex search tasks might have elicited different behaviours. Thus, further studies are needed to explore how task variation could impact user interactions with GenAI-enhanced web search.

A second caveat is that there are numerous design choices that could significantly influence patterns of interaction. The interface tested in our study relied on users to proactively engage with the chat to determine its utility. However, as Frummet and colleagues have shown, an interface that actively suggests functions or prompts follow-up actions in response to web search queries could lead to very different user behaviours and outcomes \cite{frummet2024cooking}. Similarly, the absence of source citations in the tested interface may have influenced verification behaviours, in contrast to interfaces like Perplexity\footnote{https://www.perplexity.ai/}, which now provides linked sources.

As a concluding note, our findings raise several open questions that warrant further investigation:
\begin{itemize}
    \item Are there specific ways that users can structure questions that increase the likelihood of obtaining accurate answers from Generative AI systems?
    \item What behavioural strategies can enhance successful verification, particularly in tasks where correct answers are more difficult to verify?
    \item How can information literacy techniques be adapted or developed to better support users in effectively navigating these emerging AI tools?
\end{itemize}

Understanding the answers to these questions will help refine both GenAI systems and user strategies with these, ultimately leading to more effective information-seeking and verification processes.  Further research is needed such that we can better understand how to leverage these new technologies to address our information needs.

\bibliographystyle{ACM-Reference-Format}
\bibliography{bibliography}

\end{document}